\documentstyle[preprint,aps]{revtex}
\begin{document}
\draft

\title{Cold Strangelets Formation with Finite Size Effects\\in High Energy
	Heavy-Ion Collisions
	\footnote{\it Partly supported by the National Natural 
	Science Foundation of China}}
\author{Y.B. He$^{1,2}$\footnote{E-mail: hyb@ccastb.ccast.ac.cn},
	W.Q. Chao$^{1,3,4}$, C.S. Gao$^{1,2,4}$,
 	and X.Q. Li$^{1,4,5}$}
\address{1. China Center of Advanced Science and Technology (World Laboratory),
	\\P.O. Box 8730, Beijing 100080,  China\\
	2. Department of Physics, Peking University,
         Beijing 100871, China\\
	3. Institute of High Energy Physics, Academia Sinica, P.O. Box
	918(4),	Beijing 100039, China\\
	4. Institute of Theoretical Physics, Academia Sinica, P.O. Box 
	2735, Beijing 100080, China\\
	5. Department of Physics, Nankai University, Tianjin 300071, China}
\date{\today}
\maketitle
\begin{abstract}
We have studied the phase diagram and evolution of a strangelet in equilibrium
with a finite hadronic gas. Significant finite size
modifications of the phase diagram are found and their
parameter dependences are studied. With the inclusion of finite size effects 
we have also been able to obtain
the detailed properties of the cold 
strangelet emerging in the final stage of the isentropic expansion of a 
finite strange fireball in high energy heavy-ion collisions. 
\end{abstract}
\narrowtext

\newpage
\section{Introduction}
It was speculated by Witten \cite{wit84}
that some of the dark matter in the universe might possibly
exist in the form of strange quark matter (SQM), which 
consists approximately of the same amount of up, 
down and strange quarks, and 
might be formed after the big bang when the universe
underwent the quark to hadron phase transition. 
Another place that people can expect to find SQM is in the ``little bang'',
{\it i.e.} high energy heavy-ion collisions. 
It was supposed \cite{liu84,gre87} that 
if a hot quark-gluon plasma
(QGP) is formed after the collisions, the cooling process of the QGP might 
result in droplets of SQM, so called strangelets, which could serve as an
unambiguous signature for the QGP formation. In fact, 
some heavy-ion experiments \cite{bar90}
are searching or planning to search for strangelets,
hence  reliable theoretical calculations of strangelets 
formation in heavy-ion collisions are in urgent need. 

Recently Lee and Heinz \cite{lee93}
has presented a comprehensive study of the phase structure
of a bulk strange quark matter in equilibrium with a bulk 
hadronic gas. Assuming 
Gibbs phase equilibrium and an isentropic expansion of the hot fireball,
they found under some conditions strange quark matter can actually survive
the hadronization and cooling process, giving rise to strangelet formation.
On the other hand, taking into account the particle evaporation from the
surface of the fireball, some rate 
calculations \cite{gre87,barz90}
have been carried out to study the distillation and survival
of strangelets in heavy-ion collisions.
In this work we study finite size effects on strangelet formation,
an important aspect which has been neglected in the above mentioned 
calculations.
 
It has been shown that finite size effects may strongly destabilize 
small strangelets at zero temperature \cite{mad93}, and significantly 
modify the properties of strangelets at finite temperature \cite{he96}. 
Finite size should be important as well in the study of the possible 
formation of
strangelets, which are expected to be produced 
small in heavy-ion collisions \cite{liu84,gre87}.
The present paper  follows the work of Lee and Heinz \cite{lee93}
to study the phase diagram
and evolution of strangelets in heavy-ion collisions, while 
extending it to include 
the essential finite size effects.
We find that there are indeed significant finite size modifications
of the phase diagram and evolution of strangelets.
A realistic calculation that attempts to provide reliable information for
the strangelets searches should therefore take finite size effects into account.

To study SQM in the context of the MIT bag model we have 
generally two ways to account for finite size effects. One employs 
the shell model, namely solving the Dirac equation with the bag boundary 
conditions and populating the 
energy levels of the bag \cite{far84,gre87,gil93}. However, shell model 
calculations are very tedious even at zero temperature, and become extremely
difficult when one tries to study the phase diagram and evolution. 
Another way to introduce finite size effects of strangelets is to 
explore a continuous density of states within the framework of multiple
reflection expansion \cite{bal70}, which leads to the liquid drop model
\cite{mad93,jen95}. 
As shown by Madsen \cite{mad93}, the liquid drop model reproduces the overall 
structure of shell model results for strangelets, and allows much easier
calculations than shell model. 
In this work we shall use the liquid drop model to incorporate 
finite size effects
into the studies of phase diagram and evolution of strangelets.

This paper is organized as follows. In Sect.~\ref{eos} we give a description 
of equations of state for the QGP and hadronic phase, with an emphasis on
the inclusion of finite size effects in the QGP phase.
In Sect.~\ref{pd} we study the phase diagram of a strangelet in equilibrium
with a finite hadronic phase, assuming the conservation of 
total baryon number and strangeness in the system. It is shown that the 
inclusion
of finite size effects can modify the phase diagram in a significant way.
The isentropic expansion of the system and possible formation of cold 
strangelets are explored in Sect.~\ref{fcs}. We find that in contrast with 
bulk case the inclusion of 
finite size effects will give even more 
restrictive conditions under which dynamical formation of a cold
strangelet is possible.
We summarize our work in Sect.~\ref{summary}. 

\section{Equation of State}
\label{eos}
\subsection{The QGP Phase}
We consider the QGP phase as a gas of free massless 
up and down quarks ($m_u=m_d=0$), massive strange quarks ($m_s=150$~MeV), their 
antiquarks and gluons confined in an MIT bag. 
In the multiple reflection expansion of 
Balian and Bloch \cite{bal70} the density of states for particle species $i$
present in the bag is given by,
\begin{eqnarray}
{dN_i \over dk}=g_i\left\{ {1 \over 2\pi^2} k^2 V^Q + f_S^{(i)}
\left({m_i \over k}\right)kS^Q
+f_C^{(i)}\left({m_i \over k}\right)C^Q+...\right\}.
\end{eqnarray}
For spherical QGP droplets $V^Q=4\pi (R^Q)^3/3$ is the volume of the bag,
$S^Q=4\pi (R^Q)^2$ is the surface area, and $C^Q=8\pi R^Q$ 
is the extrinsic curvature
of the bag surface. The factor $g_i$ is the statistical weight (6 for quarks
and antiquarks, and 16 for gluons). 

The functions $f_S^{(i)}$ and $f_C^{(i)}$ are surface and curvature terms for 
particle species $i$.
The surface term for quarks was given by 
Berger and Jaffe \cite{ber87} as
\begin{eqnarray}
f_S^{(q)}\left({m_q \over k}\right)=-{1 \over 8\pi}\left\{1-{2 \over \pi}
\arctan {k \over m_q}\right\}.
\end{eqnarray}
In particular, the surface term for massless quarks and gluons is zero.
The curvature term for gluons is 
$f_C^{(g)}=-1/6 \pi^2$ \cite{bal70a}, while for massless quarks
$f_C^{(q)}(0)=-1/24 \pi^2$ \cite{elz86}.
It has been shown by Madsen \cite{mad93} that the following ansatz works 
for the curvature term for massive quarks:
\begin{eqnarray}
f_C^{(q)}\left({m_q \over k}\right)={1 \over 12\pi^2}\left\{1-{3k \over 2m_q}
\left({\pi \over 2}-\arctan{k \over m_q}\right)\right\}.
\end{eqnarray}

The thermodynamic potential of the QGP phase can be derived from,
\begin{eqnarray}
\Omega = \sum_i \Omega_i + BV^Q,
\end{eqnarray}
where $B$ is the bag constant, and the thermodynamic potential of 
particle species $i$ can be written as
\begin{eqnarray}
\Omega_i & = & \mp T \int_0^{\infty} dk {dN_i \over dk} \ln[1 \pm 
\exp(-(E_i (k)-\mu_i)/T)]
\nonumber\\
& = & \Omega_{i,V}V^Q+\Omega_{i,S}S^Q+\Omega_{i,C}C^Q,
\end{eqnarray}
with the upper sign for fermions and the lower for bosons.
Here $\mu_i$ is the chemical potential. We have assumed isospin symmetry so that
$\mu_u=\mu_d=\mu_q$, and there are two independent chemical potentials
in the QGP phase, namely $\mu_q$ and $\mu_s$. To get an impression of the 
finite size effects introduced here, we have plotted in Fig.~1 the temperature
dependence of individual surface and curvature contributions of particle
species to the thermodynamic potential. One sees from Fig.~1 that they are all 
increasing functions of the temperature.
For a baryon and strangeness free QGP droplet
({\it i.e.} $\mu_q=\mu_s=0$) the surface and curvature contributions vanish
at zero temperature (Fig.~1(a) and (b)), while for nonvanishing chemical 
potentials there are finite contributions
even at zero temperature. We see that significant finite size effects can be
given by an approach of multiple reflection expansion, which can be 
relevant in the study of QGP phase structure. 

After the construction of the thermodynamic potential, 
we can readily obtain
the thermodynamic quantities of the system as follows.
The number of quarks can be derived from
\begin{eqnarray}
N_i^Q=-\left({\partial \Omega_i \over \partial \mu_i}\right)_{T,V^Q},
\end{eqnarray}
and 
\begin{eqnarray}
N_b^Q={1 \over 3} \sum_i N_i
\end{eqnarray}
gives the total baryon number of the QGP droplet.
The total entropy of the QGP phase is
\begin{eqnarray}
S^Q=-\left({\partial \Omega \over \partial T}\right)_{V^Q,\mu_i}.
\end{eqnarray}

\subsection{The Hadronic Phase}
Now we consider the equation of state for the hadronic phase.
We model the hadronic phase as a 
mixture of Bose-Einstein and Fermi-Dirac gases
of mesons $\pi$, $K$, $\eta$, baryons $N$, $\Lambda$, $\Sigma$, $\Xi$,
$\Omega$, $\Delta$ and their anti-particles \cite{hei86,jen95}.
We will show below that using this limited particle
spectrum gives practically identical results compared with the case 
in which a spectrum of all hadrons up to $M\sim$ 2~GeV is used, thus sufficient
for the study of the phase diagram.
We can write down the pressure, particle number, and energy 
density expressions for 
pointlike hadron species $i$ as follows,
\begin{eqnarray}
P_i^{pt} & = & {g_i \over 6\pi^2}\int_0^\infty dk{k^4 \over E_i(k)}
{1 \over e^{(E_i(k)-\mu_i)/T}\pm 1},
\\
N_i^{pt} & = & {g_i V^H \over 2\pi^2}\int_0^\infty dk {k^2 \over 
e^{(E_i(k)-\mu_i)/T}\pm 1},
\\
\varepsilon_i^{pt} &=& {g_i \over 2\pi^2}\int_0^\infty dk {k^2\varepsilon_i(k)
\over e^{(E_i(k)-\mu_i)/T}\pm 1},
\end{eqnarray}
where $g_i=$(spin)$\times$(isospin) is the degeneracy factor for each hadron,
the $\pm$ signs correspond to Fermi-Dirac and Bose-Eistein statistics,
$\mu_i$ is the chemical potential of hadron species $i$ and given by the 
net numbers of light quark and strange quark, $\Delta N_q^i$, 
$\Delta N_s^i$, as
\begin{eqnarray} 
\mu_i=\Delta N_q^i ~ \mu_q^H+\Delta N_s^i ~ \mu_s^H,
\end{eqnarray}
if chemical equilibriums are assumed.

The thermodynamic quantities for the hadronic gas are corrected by a Hagedorn
eigenvolume factor, $(1+\varepsilon^{pt}/4B)^{-1}$,
to take into account the repulsive interactions between
the hadrons \cite{hag80,hei86}, 
which is essential for the existence of phase transition in the
considered model, 
\begin{eqnarray}
P^H &=& {1\over 1+\varepsilon^{pt}/4B}\sum_i P_i^{pt},
\\
N_b^H &=& {1\over 1+\varepsilon^{pt}/4B}\sum_i b_i N_i^{pt},
\\
N_s^H &=& {1\over 1+\varepsilon^{pt}/4B}\sum_i s_i N_i^{pt},
\end{eqnarray}
where $\varepsilon^{pt}=\sum_i \varepsilon_i^{pt}$,
and $b_i$, $s_i$ are the baryon number, strange valence quark number of 
hadron species $i$ respectively.

We want to emphasize that a finite hadronic phase, instead of a bulk 
one \cite{jen95},
is considered in this work. In the case of a bulk QGP phase in equilibrium
with a bulk hadronic phase, the conservation of total 
net baryon number and strangeness
in the system is expressed in terms of a fixed strangeness fraction, namely
the net number of strange quarks per baryon. In our work, however, 
to include the finite size effects we have 
considered a finite QGP phase, whose volume, baryon number and
strangeness have been specified. Therefore, if the system consisting of a QGP
phase and a hadronic phase is assumed to
conserve total net baryon number and strangeness as a whole (the conservation of
total net baryon number and strangeness will be discussed furthermore below), 
the volume of the hadronic phase has to be specified as well in order to 
conserve the total net baryon number and strangeness in a consistent way.
On the other hand, we have not included surface and curvature 
contributions from hadrons in our calculations since we have not found 
appropriate expressions for them \cite{surface}. Nevertheless, there were 
expections that these contributions from hadrons may be less important 
compared with those from quarks and gluons \cite{mac75,mar91}.

\section{Phase Diagram with Finite Size Effects}
\label{pd}
We start this section with a description of our scenario. If a baryon-rich
QGP is formed in high energy heavy-ion collisions with an equal number of 
$s$ and $\bar s$ quarks, the system will suffer a strangeness enrichment
due to the early black-body radiation of more kaons 
($q\bar s$) than anti-kaons ($\bar q s$) off the fireball, and then the 
system reaches the phase coexistence region \cite{gre87}. We shall study the 
phase diagram (this section) and
further phase evolution (next section) of the resulting
system with finite net strangeness.

Following the work of Lee and 
Heinz \cite{lee93} we consider a quasistationary phase transition between
the QGP and hadronic gas, which 
assumes that the thermodynamic equilibrium time as well as the 
hadronization time are small compared to the lifetime of the mixed phase
so that a complete thermodynamic equilibrium can be held during the whole 
phase transition. 
In this work we have not taken into account the effects of 
particle evaporation off the system which may change the total net 
baryon number, strangeness content of the system \cite{gre87,barz90}. 
As a result, the total net baryon number 
in the system is kept a constant.
Furthermore, since we expect that the typical time scales in 
heavy-ion collisions are too short to 
establish a flavor equilibrium due to the weak 
interactions which can convert strange quarks to nonstrange quarks 
and vice versa, it is assumed that the total net strangeness 
in the system is conserved. 
Note that the total net baryon number and strangeness can be 
simultaneously conserved 
in a consistent way only when a finite hadronic phase is considered in 
connection with a finite QGP phase,
as we discussed in Sect.~\ref{eos}.

The Gibbs equilibrium between the QGP and hadronic phase
is given by $P^Q=P^H$ (mechanical equilibrium),
$T^Q=T^H$ (thermal equilibrium),
$\mu_q^Q=\mu_q^H$, $\mu_s^Q=\mu_s^H$ (chemical equilibrium). Taking generally
the temperatures and chemical potentials in the two phases equal, we are left
with the pressure balance between the two phases,
\begin{eqnarray}
P^Q(T,\mu_q,\mu_s;R^Q)=P^H(T,\mu_q,\mu_s).
\label{pp}
\end{eqnarray}
Note that the pressure generated by the QGP phase is a function of the QGP
droplet radius $R^Q$,
since we have included finite size effects into the equation
of state for the QGP phase.

The conservation of total 
net baryon number and strangeness in the system gives two constraints,
\begin{eqnarray}
N_b^{tot}=N_b^Q(T,\mu_q,\mu_s;R^Q)+N_b^H(T,\mu_q,\mu_s;V^H),
\label{nb}
\end{eqnarray}
and 
\begin{eqnarray}
N_s^{tot}=N_s^Q(T,\mu_q,\mu_s;R^Q)+N_s^H(T,\mu_q,\mu_s;V^H),
\label{ns}
\end{eqnarray}
where the volume of the hadronic phase is given by
\begin{eqnarray}
V^H=V^{tot}-V^Q.
\label{vh}
\end{eqnarray}

The phase diagram of the system is determined by Eqs.~(\ref{pp})$-$(\ref{vh}),
and shown in Fig.~2(a) for parameters $B^{1/4}=180$~MeV, $m_s=150$~MeV,
and a fixed strangeness fraction (namely net number of 
strange  quarks per baryon) $f_s=0.5$.
Full lines in Fig.~2(a) illustrate our results 
for a finite system with total net baryon number $N_b^{tot}=100$
and strange quark number $N_s^{tot}=50$.
To be more specific, the full line separating the pure QGP phase and mixed 
phase (MP) is obtained by setting $V^H=0$ in Eqs.~(\ref{nb})
and (\ref{ns}), and solving Eqs.~(\ref{pp})$-$(\ref{ns}) for ($\mu_q$,
$\mu_s$, $R^Q$) at a given $T$, $N_b^{tot}$ and $N_s^{tot}$.
On the other hand, drawing the boundary between the mixed phase (MP) 
and hadronic gas (HG) is a little more delicate. 
For a given value of $V^{tot}$, we
solve Eqs.~(\ref{pp})$-$(\ref{vh}) for ($\mu_q$, $\mu_s$, $R^Q$, $V^H$)
at fixed $T$, $N_b^{tot}$ and $N_s^{tot}$.
We then increase the value of $V^{tot}$ until $V^{tot}$
is so large that Eqs.~(\ref{pp})$-$(\ref{vh}) have no solutions for 
($\mu_q$, $\mu_s$, $R^Q$, $V^H$), which implies that the QGP and hadronic phase
can no longer coexist in equilibrium, and thus we have reached the boundary
between the mixed phase (MP) and hadronic gas (HG). We observe that at the 
MP$-$HG boundary the QGP droplet may have non-vanishing
baryon number and strangeness, which is of course a consequence of 
the inclusion of finite size effects. 
To check the consistency of this solution 
procedure, we have also tried to decrease the
value of $V^{tot}$ while solving  Eqs.~(\ref{pp})$-$(\ref{vh}) 
until no solutions exist, and found that 
we exactly arrived at the QGP$-$MP boundary,
which we have obtained above.

In comparison with the above case of a finite system, the dashed 
curves in Fig.~2(a) show the phase 
diagram of a bulk QGP in equilibrium with a bulk 
hadronic phase, {\it i.e.} without finite size effects, which is 
identical to the results
of Lee and Heinz in Fig.~6(c) of Ref.~\cite{lee93}.
One sees from Fig.~2(a) that the hadronic phase grows significantly with the 
inclusion of finite size effects, 
while the QGP$-$MP boundary is little affected by the finite size corrections.
We shall show in next section that this growth of the hadronic 
phase will greatly reduce
the survival probabilities of cold strangelets which we wish to find in 
heavy-ion collisions.

The dotted lines in Fig.~2(a) is plotted for $N_b^{tot}=50$, $N_s^{tot}=25$,
to show how the finite size modifications of phase diagram vary with the given
$N_b^{tot}$ and $N_s^{tot}$, in comparison with the bulk case (dashed lines)
and the case of $N_b^{tot}=100$ and $N_s^{tot}=50$ (full lines). 
It is of no surprise that with smaller $N_b^{tot}$ and $N_s^{tot}$ the smaller
system has its phase diagram modified more prominently
by the finite size effects.  

To see how the modifications of phase diagram due to the inclusion of 
finite size effects are affected by the bag constant $B$, we have plotted
in Fig.~2(b) and 2(c) the phase diagrams for $B^{1/4}=145$~MeV and 235~MeV, 
$f_s=0.5$, with all dashed lines for a bulk system, 
and all full lines for a finite system 
with total net baryon number $N_b^{tot}=100$,
strange quark number $N_s^{tot}=50$. 
We see that finite size modifications are more significant for larger
bag constants. This can be understood roughly that for  given $N_b^{tot}$,
$N_s^{tot}$, and $T$, larger bag constants 
generally correspond to more compact and smaller QGP droplets in equilibrium 
with a hadronic gas, and thus more important finite size effects. 
However, we want to point out that the full 
line in Fig.~2(c) separating the mixed phase (MP) and hadronic gas (HG) is 
somehow out of control at regions of high baryon number density $\rho_b$. 
This difficulty may come form two origins. First, at regions of high 
baryon number density the chemical potentials may become very large and 
lead to some numerical uncertainty in the relevant Bose-Einstein 
integrations, which can be solved by further careful examinations. 
Another possibility is that 
at these regions the radius $R^Q$ of QGP 
droplet in equilibrium with a finite
hadronic gas tends to become so small ($\alt 1$~fm)
that the multiple reflection expansion
in powers of $1/R^Q$, which we have adopted to introduce finite size effects,
could be somewhat dangerous, as previously warned by Madsen \cite{mad93}. 
Fortunately, in our calculations 
this kind of disease sets in generally for very large bag 
constants.
As it is expected
that for sufficiently large bag constant strangelets can neither be stable nor
metastable even at zero temperature and the formation of cold strangelets in 
heavy-ion collisions is very unlikely \cite{gre87,barz90}, 
therefore this problem will not become 
serious if we are only interested in the cases in which the formation of a 
cold strangelet can be possible.

\section{Formation of Cold Strangelets}
\label{fcs}
In this section we turn to study the evolution of the system to see how a cold
strangelet could be formed in the final stage of evolution 
and what its characteristics look like.
While a full dynamical investigation of the fireball evolution
is beyond the scope of this
work, we assume that the system expands adiabatically, conserving 
its total entropy during the whole evolution \cite{hei86,sto80}. 
Note again that we have not included the process of surface evaporation which
may change the entropy, net baryon number and strangeness of the system.
Therefore what we are exploring is a system that experiences 
a smooth hydrodynamic expansion, conserving the total entropy
$S^{tot}$,
net baryon number $N_b^{tot}$, and strangeness $N_s^{tot}$ \cite{gre87}.
As we have already described in Sect.~\ref{pd}, the system considered here
may have nonvanishing net strangeness due to the early surface radiation
of hadrons after the collisions. 

We first look at a QGP fireball which is expanding isentropically and cooling
down until it reaches the boundary of mixed phase.
The isentropic expansion trajectories in the pure QGP phase are determined by
solving 
\begin{eqnarray}
N_b^{tot}=N_b^Q(T,\mu_q,\mu_s;R^Q),
\\ 
N_s^{tot}=N_s^Q(T,\mu_q,\mu_s;R^Q),
\end{eqnarray} 
and 
\begin{eqnarray}
S^{tot}=S^Q(T,\mu_q,\mu_s;R^Q)
\end{eqnarray}
for $(T,\mu_q,\mu_s)$  
at fixed $N_b^{tot}$, $N_s^{tot}$, $S^{tot}$ and various $R^Q$.
Increasing the value of $R^Q$ until the pressure balance Eq.~(\ref{pp})
is possible, we will arrive at the boundary between the QGP and
mixed phase, which we have obtained in the phase diagram given in last section.

Afterwards the isentropic expansion trajectories go into 
the mixed phase and are
given by Eqs.~(\ref{pp})$-$(\ref{vh}), together with  
\begin{eqnarray}
S^{tot}=S^Q(T,\mu_q,\mu_s;R^Q)+S^H(T,\mu_q,\mu_s;V^H),
\label{stot}
\end{eqnarray}
which are solved for $(T,\mu_q,\mu_s,R^Q,V^H)$
at fixed $N_b^{tot}$, $N_s^{tot}$, $S^{tot}$ and various $V^{tot}$.
When we increase the value of $V^{tot}$ while solving these equations,
two phenomena may be observed. One is that these equations have no solutions
for $(T,\mu_q,\mu_s,R^Q,V^H)$ when $V^{tot}$ is large enough, which is 
found to be the very case in which the expansion 
trajectory hits the MP$-$HG boundary
at some point. This observation further confirms the consistency of our 
treatment of finite size effects. In this case the expansion trajectory
will then enter the 
hadronic phase, corresponding to a complete hadronization,
and no cold strangelet could be formed.
Another important phenomenon is that for $V^{tot}\rightarrow \infty$
Eqs.~(\ref{pp})$-$(\ref{vh}) and (\ref{stot})
can always be solved
at fixed $N_b^{tot}$, $N_s^{tot}$ and $S^{tot}$, giving $T\rightarrow 0$.
This implies that a cold strangelet could appear in the final stage of 
the evolution, which is of great interest in heavy-ion experiments.

As manifestations of the above discussions, we show in Fig.~3 
isentropic expansion trajectories of the system through the phase diagrams
for $B^{1/4}=180$~MeV, $N_b^{tot}=100$,
and $N_s^{tot}=$~0, 50, 100, 200, or strangeness fractions $f_s=$~0, 0.5, 
1.0, 2.0. Our Fig.~3 can be compared with the results of Lee and Heinz for bulk
phases given in Fig.~6 of Ref.~\cite{lee93} to see how the inclusion of finite
size effects can modify the isentropic expansion trajectories of the system. 
It can be seen in Fig.~3 that for most cases 
the expanding system will hadronize completely. 
In particular, for $f_s=0.5$ and $S^{tot}/N_b^{tot}=5$ (Fig.~3(b)), 
in contrast with the bulk case (Fig.~6(c) in Ref.~\cite{lee93})
in which the system will always stay inside the mixed phase
and give rise to cold strangelet formation, one finds that 
the inclusion of finite 
size effects tends to make the formation of cold strangelets even more 
difficult. In fact, with an impressive observation from 
Fig.~2(a) on how 
the phase diagram can be drastically modified by the finite size of the system,
we can expect that currently existing calculations of possible
formation of cold strangelets in heavy-ion collisions without finite size 
effects may have to become somehow less optimistic with 
the inclusion of finite size effects.

The inclusion of finite size effects may also affect in some way 
the strangeness separation process during the phase transition which is
an essential ingredient for the possible formation of cold 
strangelets in heavy-ion  collisions \cite{gre87}. 
Fig.~4 shows the evolution details of the
QGP droplet corresponding to the isentropic expansion
trajectory for a system with zero net strangeness and 
$S^{tot}/N_b^{tot}=5$ in Fig.~3(a). In Fig.~4(a) the system 
experiences a cooling in the 
pure QGP phase, and reheating in the mixed phase which is mainly due to 
the energy surplus when quarks in the QGP are converted into hadrons in 
the hadronic gas. Fig.~4(b) shows a smooth variation of chemical potentials 
$\mu_q$ and $\mu_s$ during the phase transition. In particular, 
$\mu_s$ remains zero in the pure QGP phase whose net strangeness is zero, but 
increase continuously in the mixed phase even though the net strangeness of
the system remains zero, indicating the occurrence of strangeness separation
which can be seen also in Fig.~4(c) from a continuous increase of the 
strangeness fraction $f_s^Q$ of the QGP droplet.
During the evolution in the mixed phase
the net baryon number $N_b^Q$ (Fig.~4(d)) and radius $R^Q$ 
(Fig.~4(e)) of the QGP droplet decrease monotonously. 
However, at the boundary between the mixed phase and hadronic gas $N_b^Q$
and $R^Q$ do not vanish, and only a rather
moderate value of $f_s^Q$ can be reached. This is quite different from the
corresponding cases for bulk phases studied previously \cite{gre87,hei87}, 
in which the QGP phase diminishes at the MP$-$HG phase boundary and 
the strangeness separation mechanism works in a much more effective way,
leading to a much larger value of $f_s^Q$.
In fact, as we have already known in the studies of phase diagrams 
that the hadronic
phase grows on account of the mixed phase with the inclusion of the finite 
size effects, we can expect that the inclusion of the finite size 
effects tends to make the strangeness separation proceed to a less extent.
Since the strangeness separation mechanism appears so important in calculations
of the possible formation of cold strangelets in heavy-ion collisions, 
we hereby feel 
it a substantial task to include finite size effects in the future 
relevant investigations.

Now we study the formation of cold strangelets in heavy-ion collisions. 
The isentropic expansion trajectories in 
Fig.~3(c) and (d) indicate that for $S^{tot}/N_b^{tot}=5$ the system will 
always stay in the mixed phase and expand infinitely to zero 
temperature and zero baryon number density, 
resulting in the
formation of a cold strangelet. As an essential consequence of the inclusion
of finite size effects, we can now obtain the properties of the surviving
strangelet. For example, we list in Table.~\ref{slet} 
the properties of the resulting cold 
strangelet corresponding to the expansion trajectories for 
$S^{tot}/N_b^{tot}$=5
and $f_s^{tot}=1.0$ (Fig.~3(c)), 2.0 (Fig.~3(d)) respectively.
Note in the case of $f_s^{tot}=1.0$ all net strangeness is in the droplet
($N_s^Q=100$), which is due to the fact that at $T\rightarrow 0$ chemical
potentials $\mu_q$, $\mu_s$ in the phase 
equilibrium configuration acquire values 
such that only nucleons but no strange baryons 
are present in the hadronic gas.  

However, the properties of cold strangelets given in Table.~\ref{slet} 
should not be taken too seriously for two reasons. First, an isentropic 
and equilibrium expansion of the fireball all the way to zero temperature 
assumed in this work is
not a realistic scenario. Effects of particle evaporations off the 
surface of the fireball can be very important for the strangelet formation 
\cite{gre87,barz90}.
At sufficiently low temperature and  density the QGP droplet may decouple
from the surrounding hadronic gas and the thermodynamic equilibriums between
them may be broken. On the other hand, the listed properties of the strangelets
rely heavily on the parameters such as bag constant $B$, strange quark mass
$m_s$, and on the given initial conditions of the system in terms of 
$N_b^{tot}$, $N_s^{tot}$ and $S^{tot}$. Nevertheless, Table.~\ref{slet} can
give us a first  impression of the characteristics of cold strangelets 
resulted from a simple dynamical evolution, and may provide clues for 
future construction of realistic dynamical evolution models. 

The isentropic expansion trajectories in Fig.~3 imply that only for 
sufficiently low $S^{tot}/N_b^{tot}$ or large $f_s$ does the system allow 
the formation of cold strangelets. Following Lee and Heinz \cite{lee93} we have 
plotted in Fig.~5 
the maximum total entropy per baryon $S^{tot}/N_b^{tot}$ permitted by the 
formation of cold strangelets versus the  strangeness fraction 
$f_s$, for two bag constants $B^{1/4}=145$ and 180~MeV, and a finite 
system with net baryon number $N_b^{tot}=100$. In comparison 
with the results for bulk phases in Fig.~8 of Ref.~\cite{lee93}, 
we see that the inclusion of finite size effects give much more 
restrictive conditions under which cold strangelets could be formed.
Even for bag constant as small as $B^{1/4}=145$~MeV, which is expected to 
be the minimum value compatible with the usual nuclear matter phenomenology,
and for strangeness fraction $f_s=3.0$,
an upper limit $S^{tot}/N_b^{tot}\alt 52$ is required to make the formation
of cold strangelets possible. 
However, these results may depend on the net baryon number $N_b^{tot}$, or 
the size of the system, and are to be made more convincing by further detailed
investigations.

\section{Summary}
\label{summary}
In the framework of multiple reflection expansion we have studied finite size
effects on the phase diagram and evolution of a strangelet in 
equilibrium with a finite hadronic gas. Our aim is to check how important the
finite size effects can be in the studies of possible formation of 
cold strangelets in high energy heavy-ion collisions.
We have shown that there are indeed significant finite size modifications of
the phase diagram, which are found to 
become more important for a system with
larger bag constant $B$ or smaller total net baryon number $N_b^{tot}$. 

Assuming an isentropic expansion of the 
system we have studied the finite size effects on the evolution of a 
strangelet. It is observed that the inclusion of finite size effects tends to 
make the strangeness separation mechanism less effective, and will
give very
restrictive conditions for the cold strangelet formation in 
heavy-ion collisions, expressed in terms of windows of
the bag constant $B$, the total
entropy per baryon $S^{tot}/N_b^{tot}$, and the strangeness fraction 
$f_s^{tot}$. 
With the inclusion of finite size effects we have been able to obtain 
quantitatively the properties of cold strangelets emerging in the final stage 
of the isentropic expansion of the system. 
 
Even though the treatment of finite size effects within the framework of 
multiple reflection expansion may fail for very small strangelets as we have
uncovered in the studies of the phase diagrams, and the isentropic
and equilibrium evolution scenario studied in this work is not a realistic
one, yet we can still conclude after this work 
that finite
size effects do play an essential role and should be included in future 
realistic investigations of the possible formation of cold strangelets in 
high energy heavy-ion collisions.
\acknowledgments
The authors would like to thank Prof. R. K. Su and Dr. S. Gao
for helpful discussions.

\newpage
\narrowtext
\begin{table}
\caption[Properties of surviving strangelets]{Properties of surviving 
strangelets for a system with 
$B^{1/4}=180$~MeV, $m_s=150$~MeV, $N_b^{tot}=100$, and $S^{tot}/N_b^{tot}=5$.
The second column describes a cold strangelet resulting from the isentropic 
expansion trajectory $S^{tot}/N_b^{tot}=5$ in Fig.~3(c) with $N_s^{tot}=100$ 
or equivalently $f_s^{tot}=1.0$, and the third column corresponds to the 
isentropic expansion trajectory $S^{tot}/N_b^{tot}=5$ in Fig.~3(d) with 
$N_s^{tot}=200$ or $f_s^{tot}=2.0$.}
\begin{tabular}{ccc}
 & $f_s^{tot}=1.0$ & $f_s^{tot}=2.0$\\ 
\tableline
$R^Q$ (fm) & 3.07 & 3.43 \\ 
$N_b^Q$& 62.7& 84.2 \\ 
$\rho_b^Q=N_b^Q/V^Q$ (fm$^{-3}$) & 0.52& 0.50 \\ 
$N_s^Q$& 100.& 170. \\ 
$f_s^Q=N_s^Q/N_b^Q$& 1.59& 2.02 \\ 
$Z/N_b^Q=(1-f_s^Q)/2$& $-0.30$ & $-0.51$ \\ 
$E^Q/N_b^Q$ (MeV)& 1184& 1250 \\ 
\end{tabular}
\label{slet}
\end{table}

\newpage
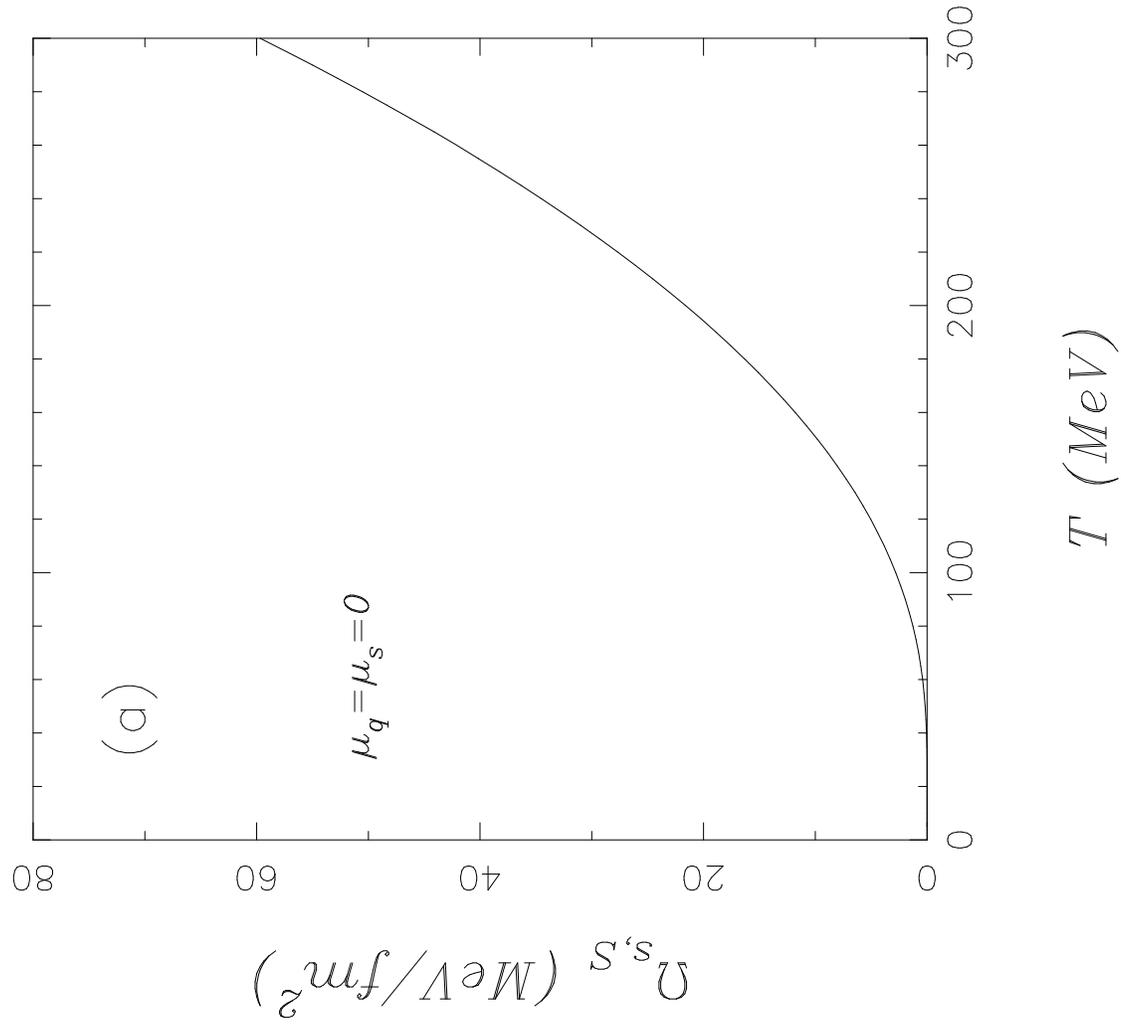
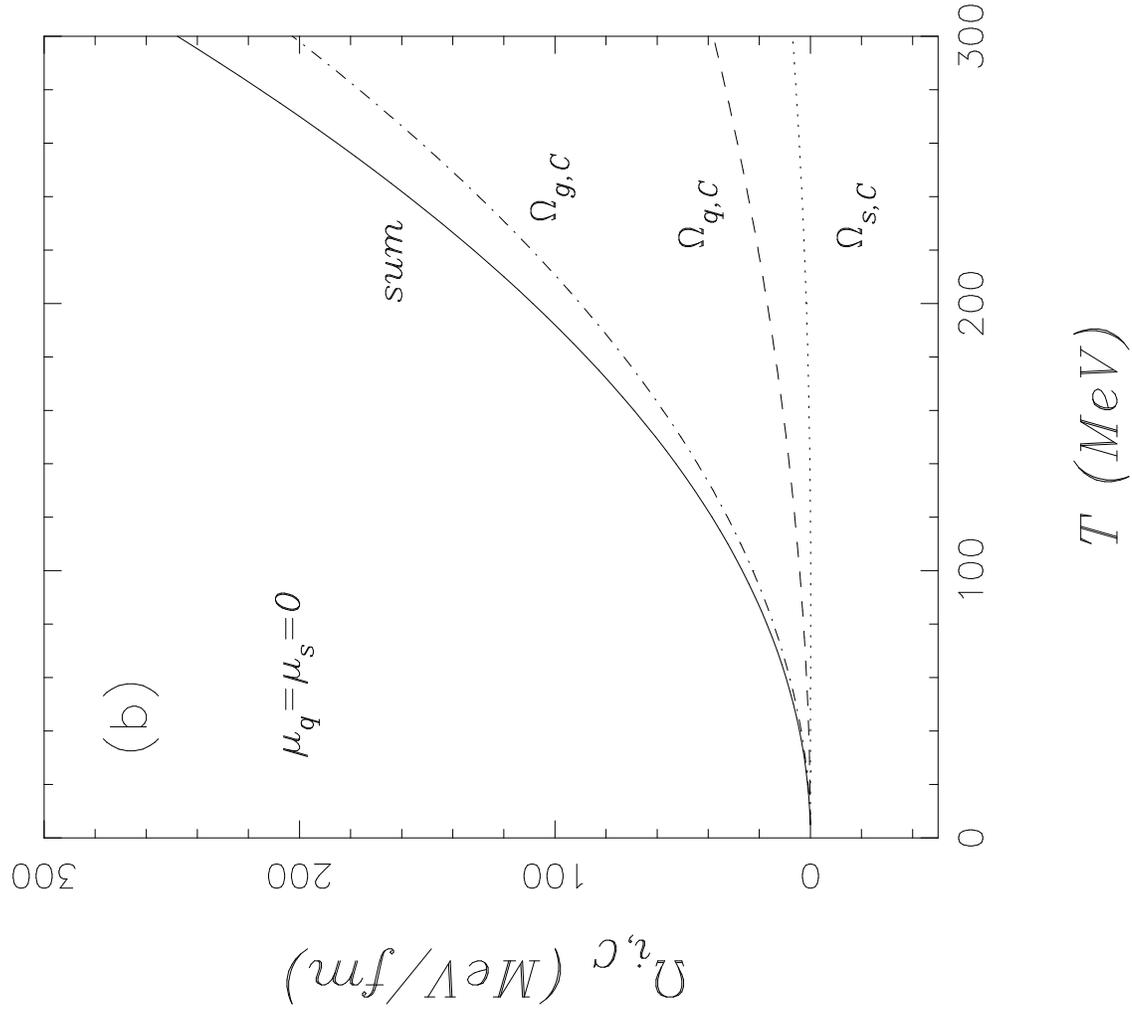
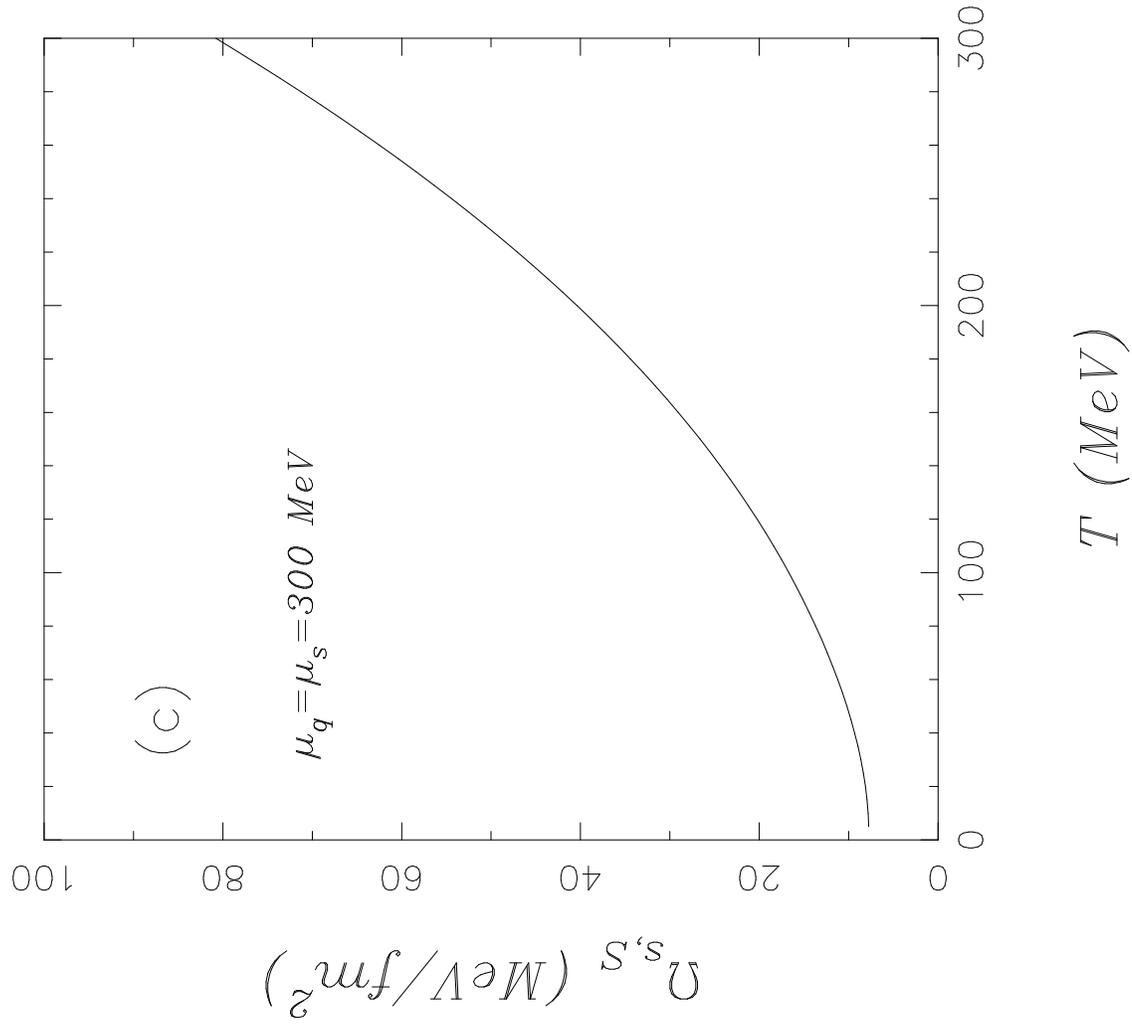
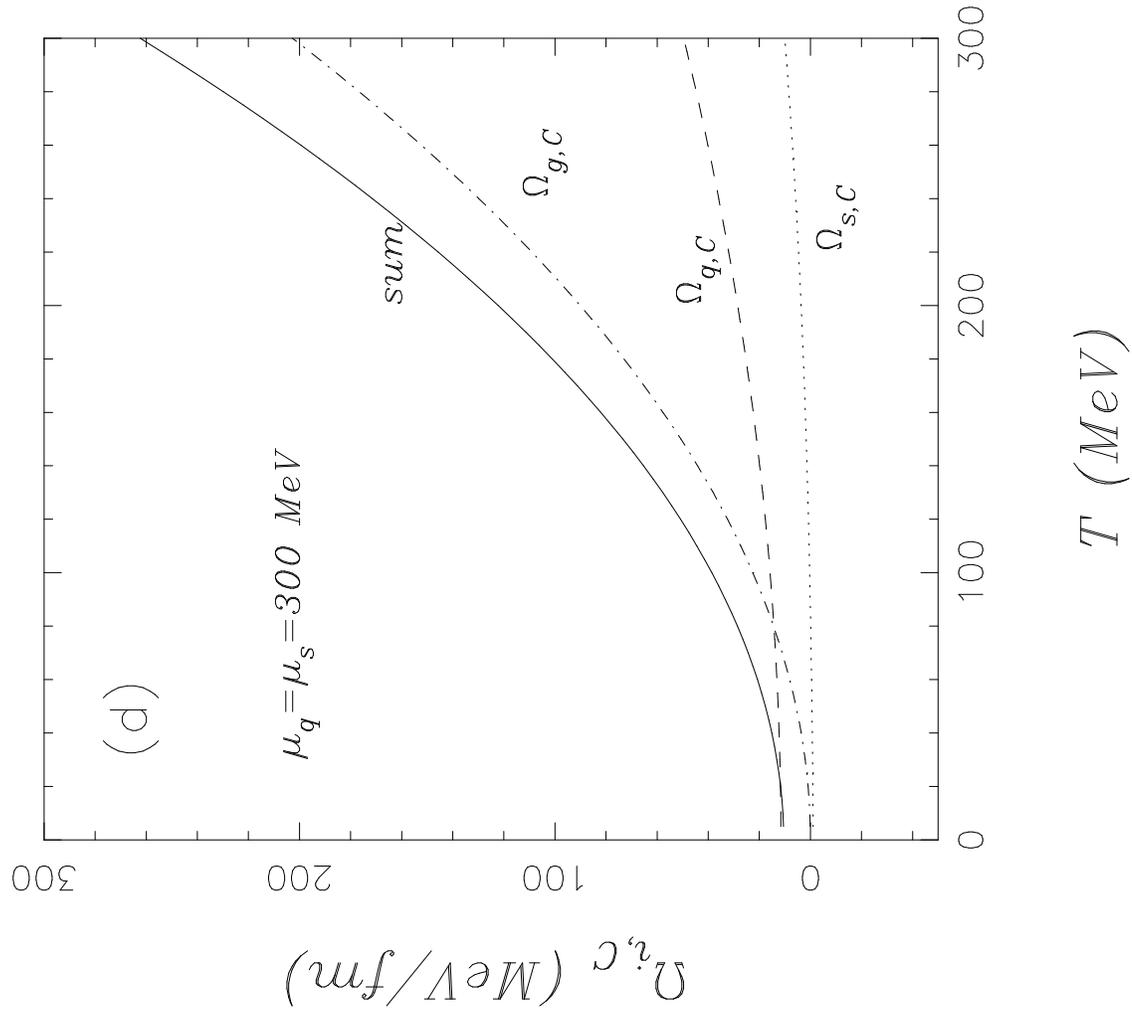
\begin{figure}
\caption{Temperature dependences of the surface and curvature contributions 
to the thermodynamic potential of light quarks (``q''),
massive strange quark (``s''), and gluon (``g'') for 
vanishing and nonvanishing chemical potentials. The ``sum'' in (b) and (d)
stands for the total curvature contributions.}
\label{fig1}
\end{figure}

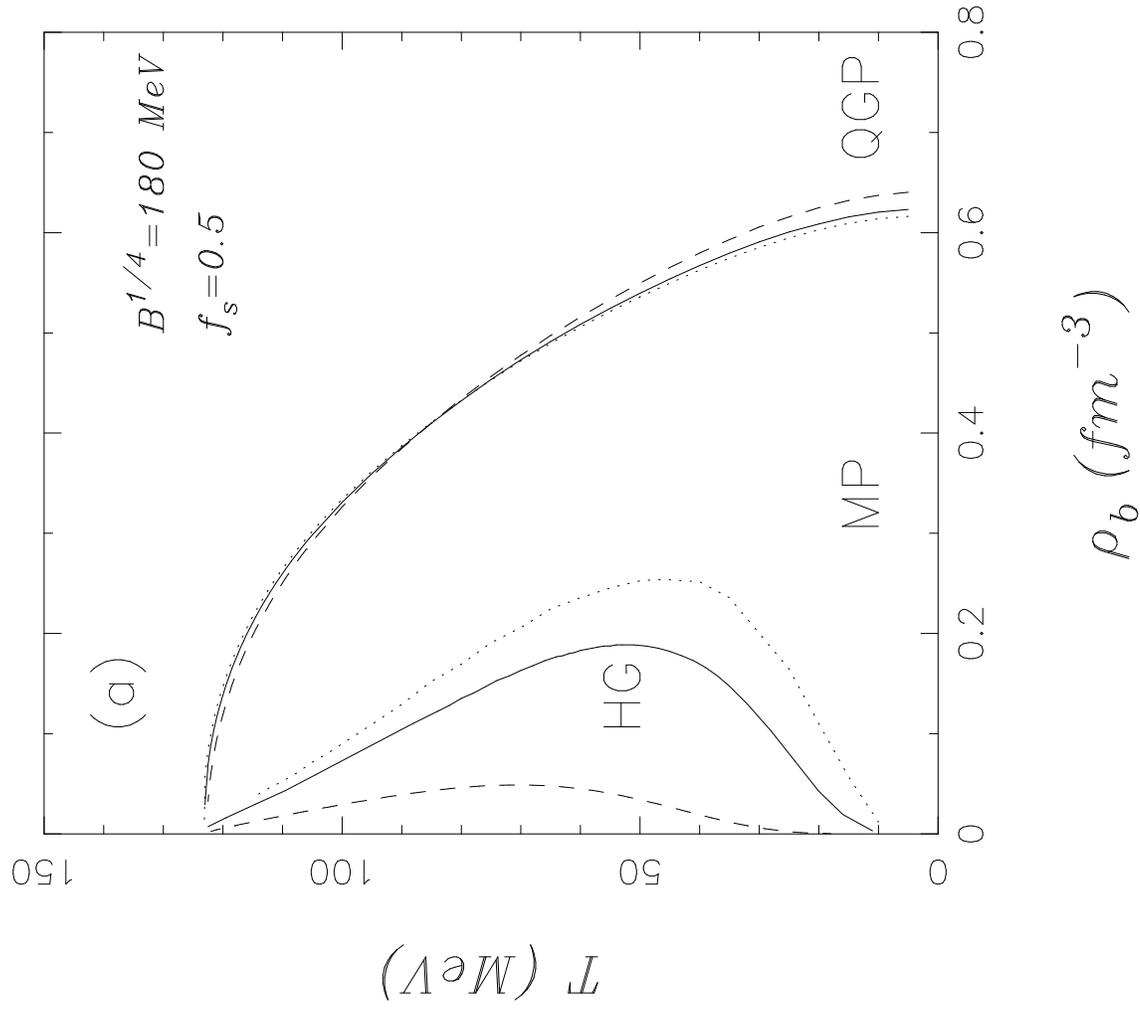
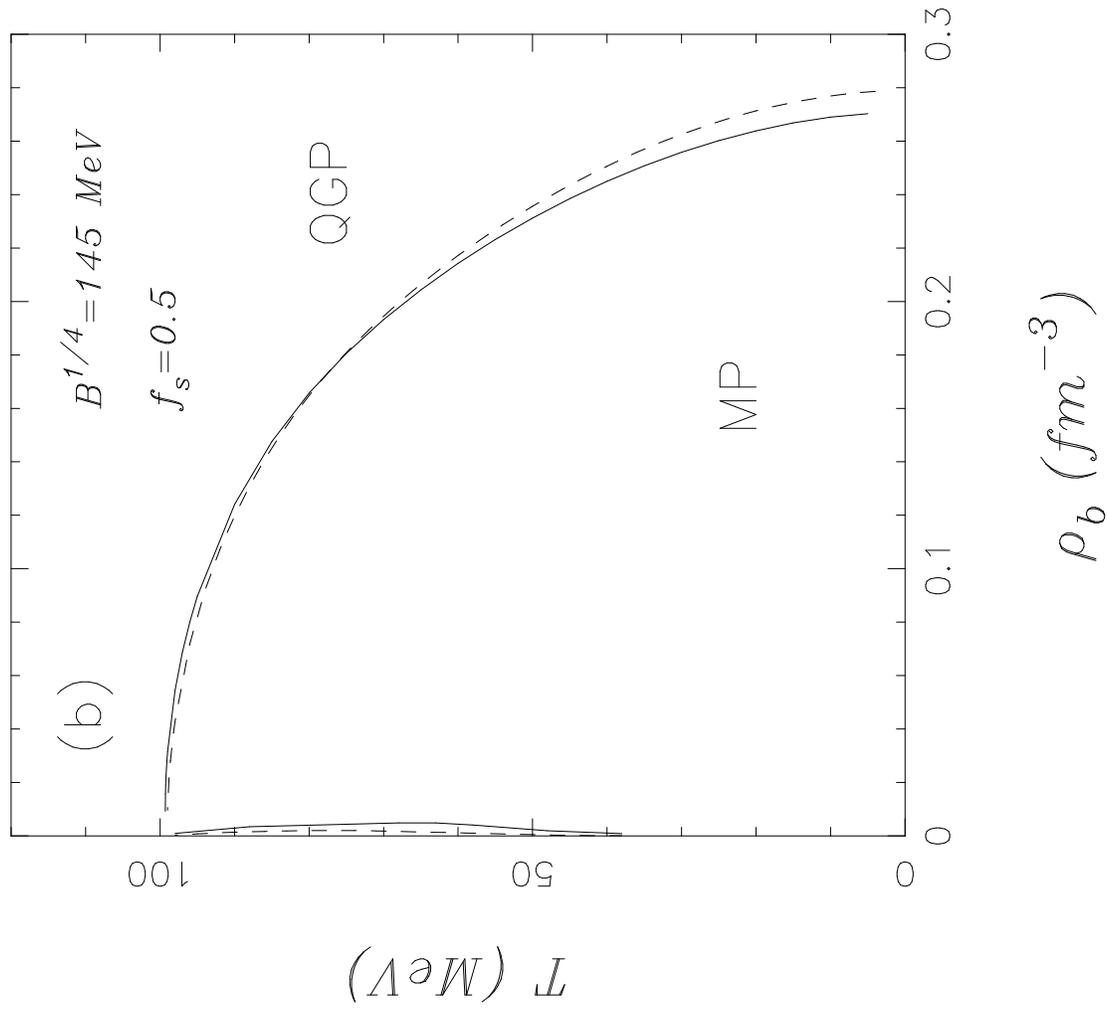
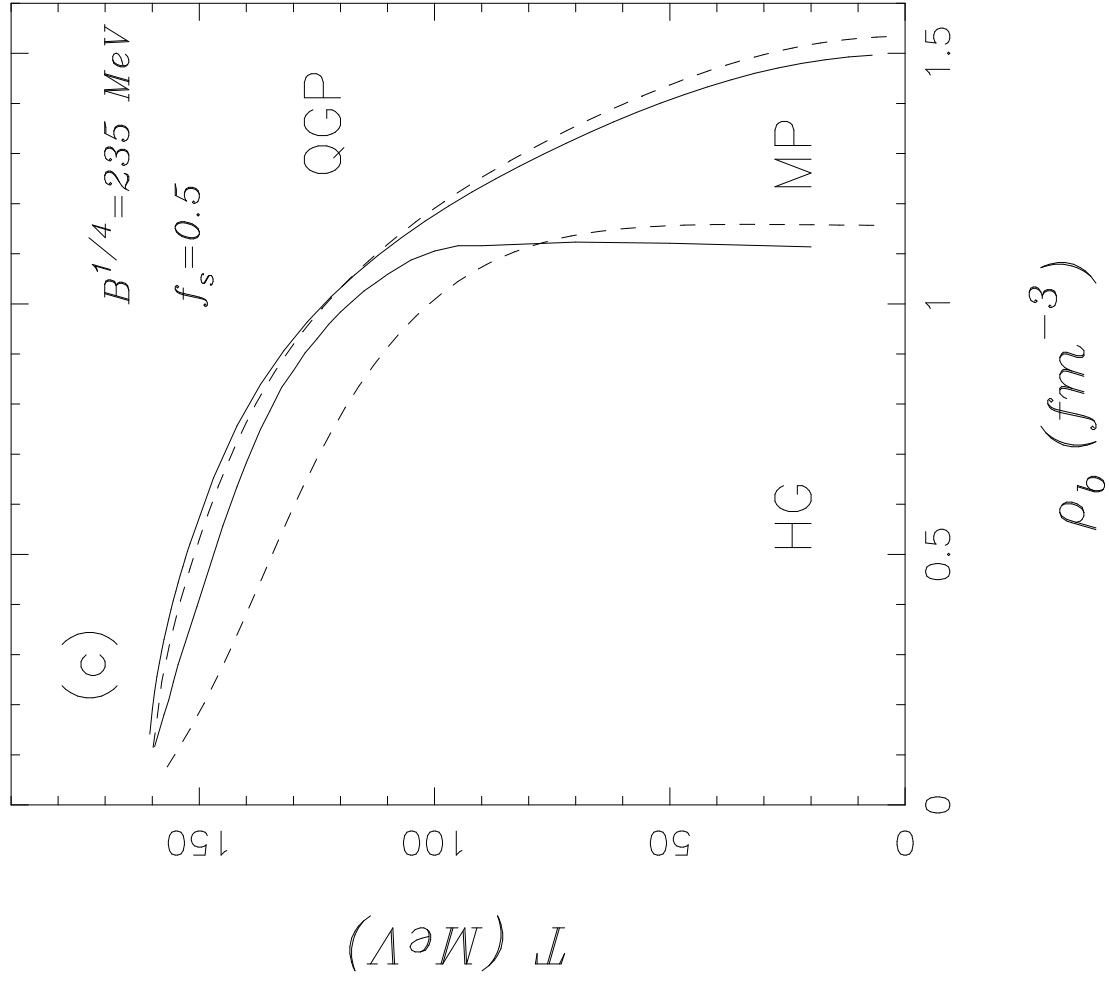
\begin{figure}
\caption{(a) Phase diagrams for a QGP phase in equilibrium with 
a hadronic gas (HG),
with bag constant $B^{1/4}=180$~MeV and fixed  strangeness fraction $f_s=0.5$:
dashed lines for a bulk QGP and a bulk hadronic gas;
full lines for a finite system with total net baryon number $N_b^{tot}=100$ 
and net strange quark number $N_s^{tot}=50$; dotted lines for $N_b^{tot}=50$
and $N_s^{tot}=25$. Note that the pure QGP phase and the hadronic gas (HG) are
separated by a finite mixed phase (MP). 
(b) Phase diagrams for a QGP phase in equilibrium with 
a hadronic gas (HG),
with $B^{1/4}=145$~MeV and fixed $f_s=0.5$:
dashed lines for a bulk QGP and a bulk hadronic gas;
full lines for a finite system with $N_b^{tot}=100$ 
and $N_s^{tot}=50$.
(c) The same as in (b) but for $B^{1/4}=235$~MeV.} 
\label{fig2}
\end{figure}

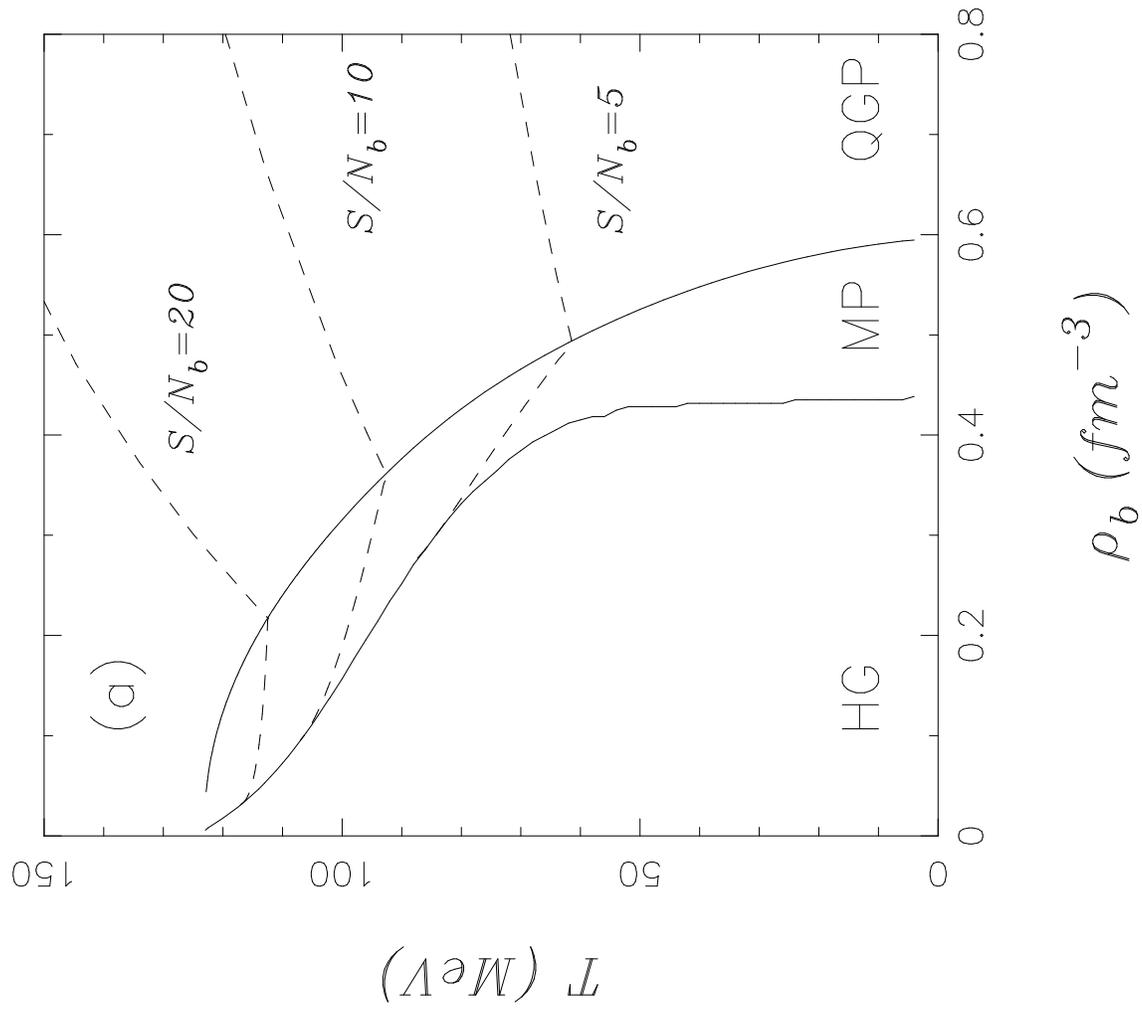
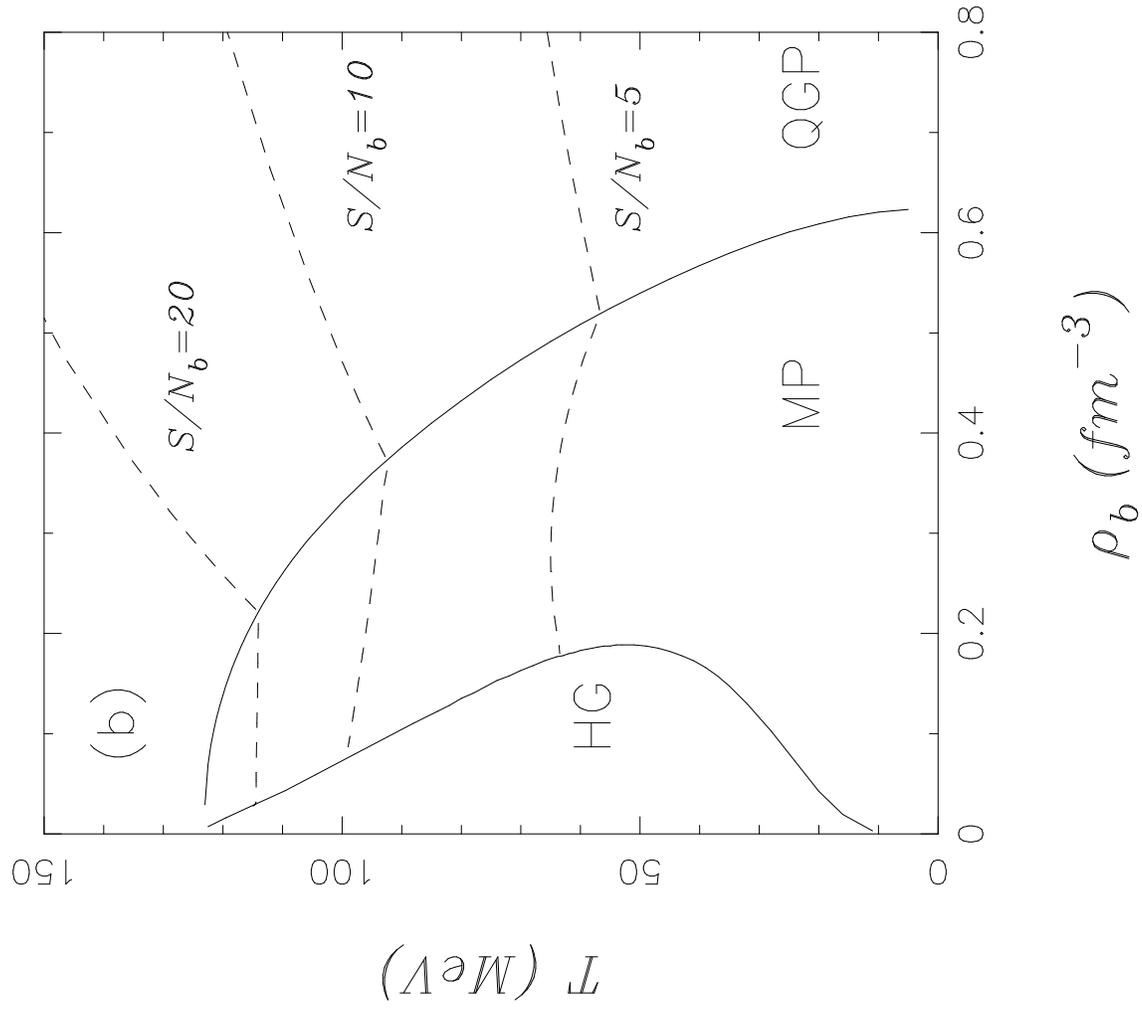
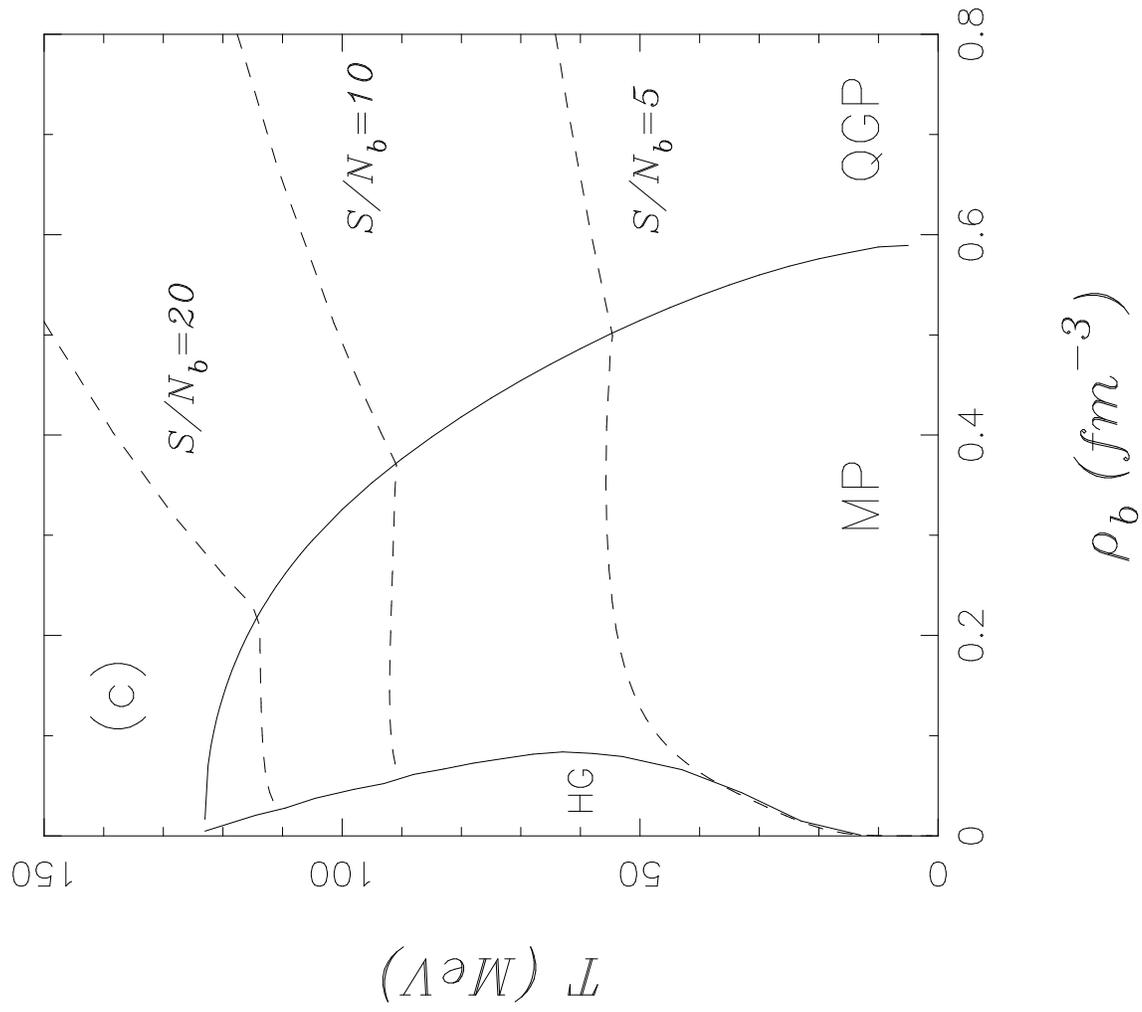
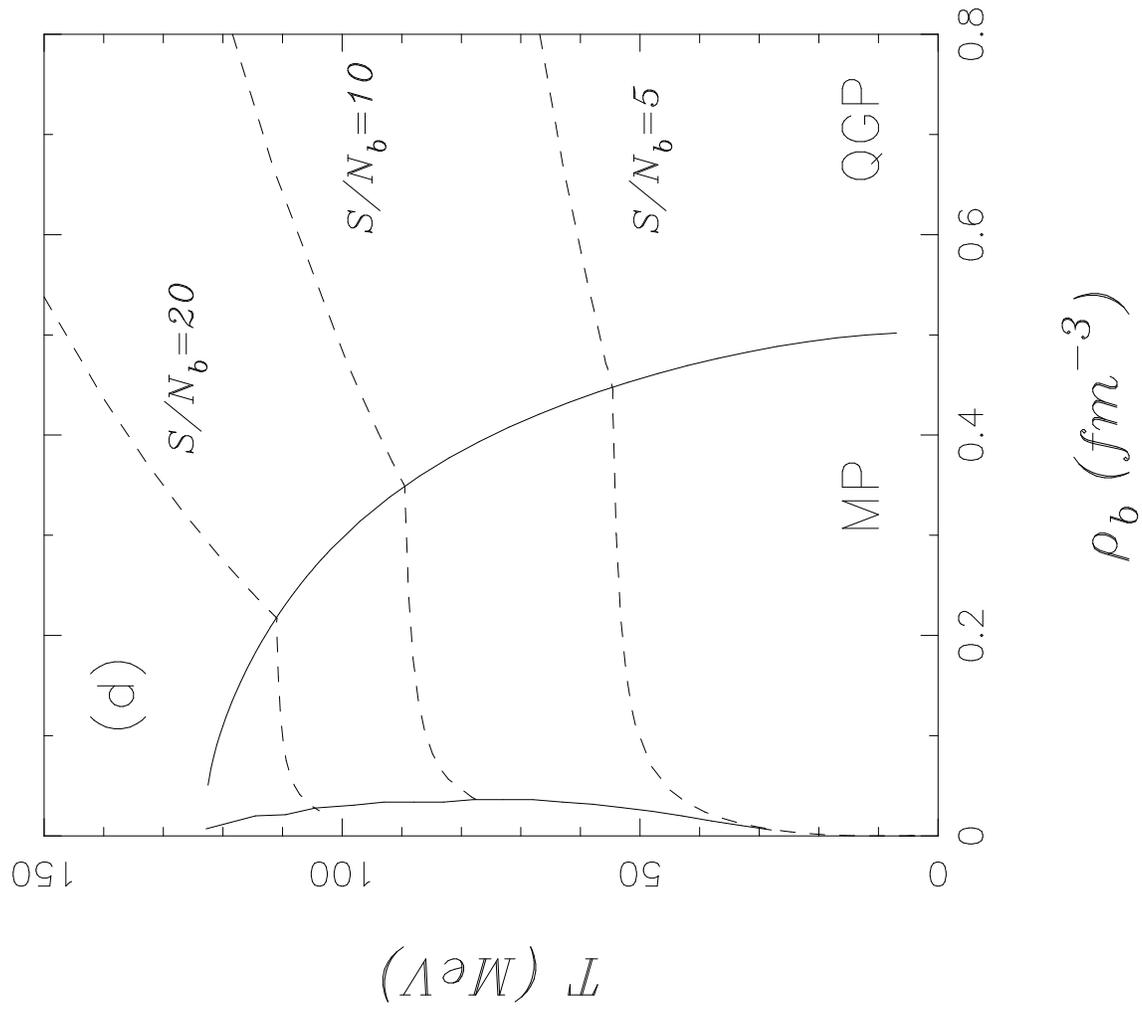
\begin{figure}
\caption{Isentropic expansion trajectories of a finite system 
with various total entropy per baryon $S^{tot}/N_b^{tot}$,
total net baryon number 
$N_b^{tot}=100$, and net strange quark number (a) $N_s^{tot}=0$; 
(b) $N_s^{tot}=50$; (c) $N_s^{tot}=100$; (d) $N_s^{tot}=200$. 
($B^{1/4}=180$~MeV, $m_s=150$~MeV)}
\label{fig3}
\end{figure}

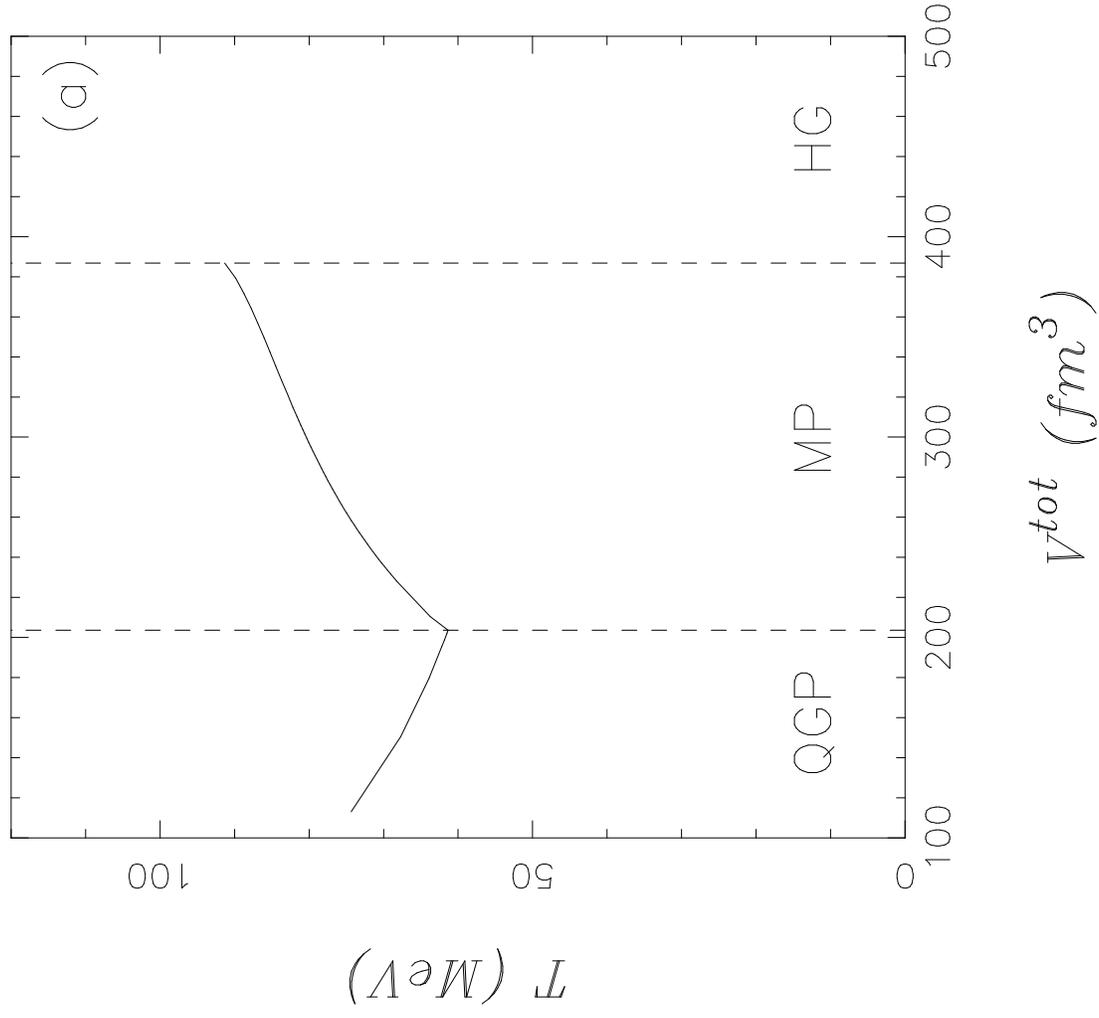
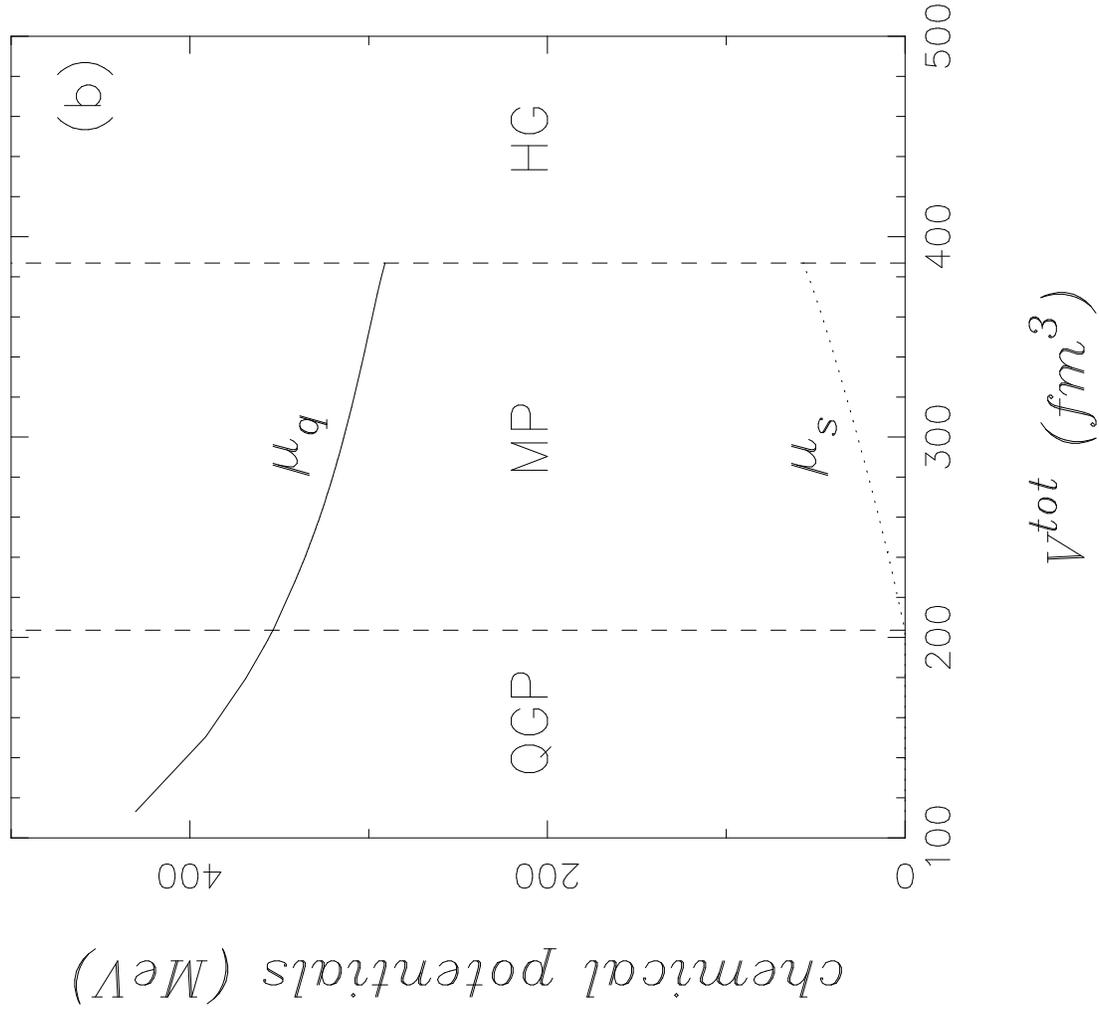
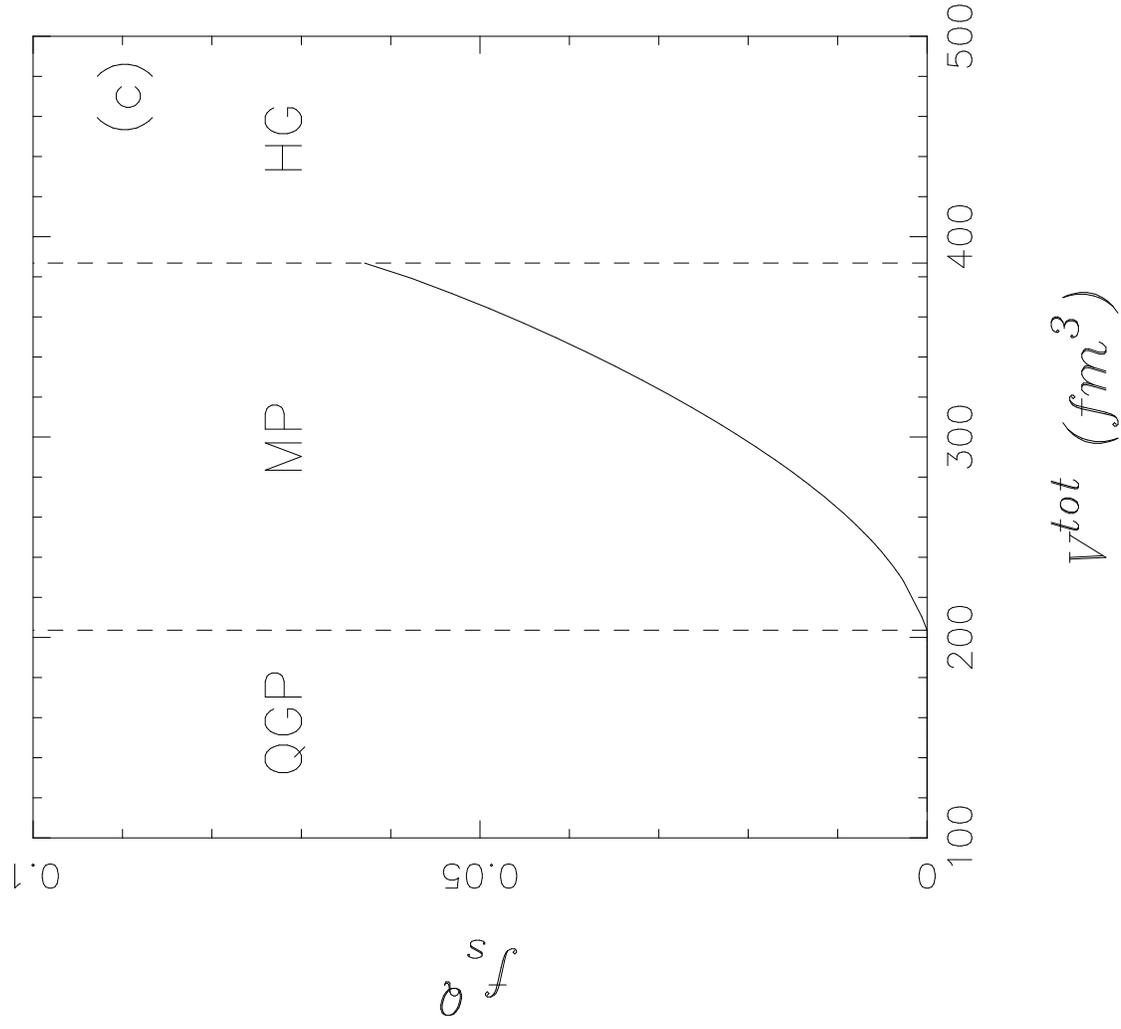
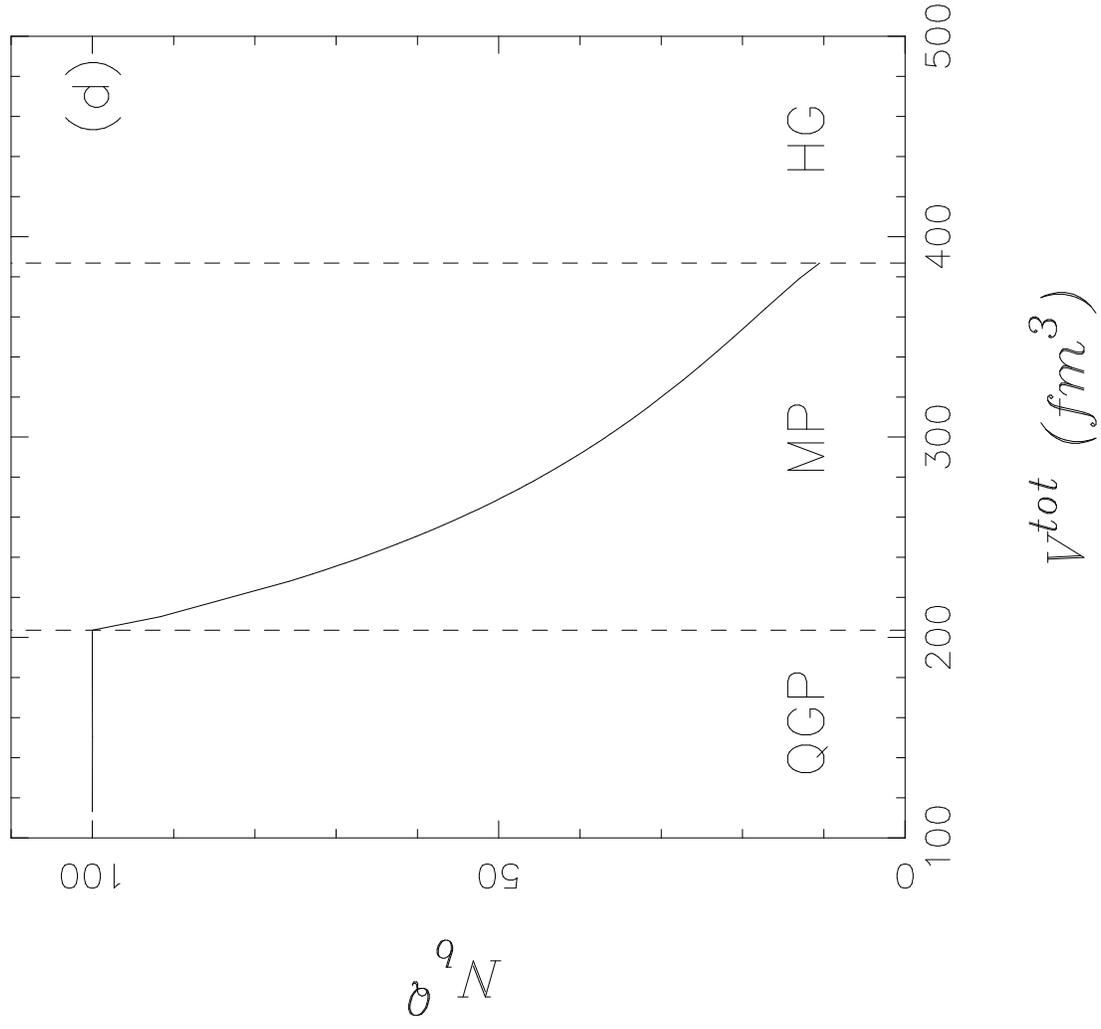
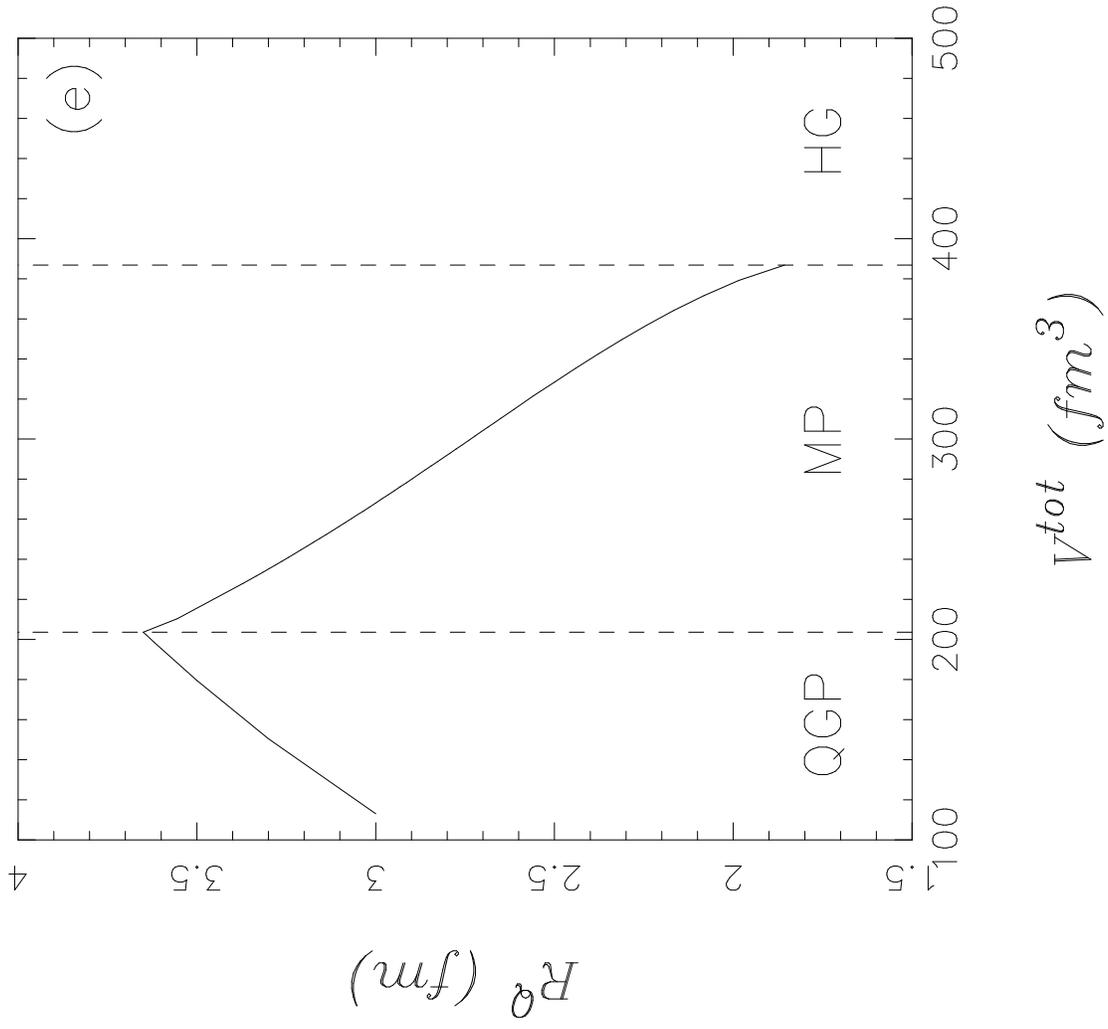
\begin{figure}
\caption{Variations of (a) temperature $T$, (b) chemical potentials $\mu_q$,
$\mu_s$, (c) strangeness fraction  $f_s^Q$, (d) net baryon
number $N_b^Q$, and (e) radius $R^Q$ of the QGP droplet during the isentropic
expansion of a system with $S^{tot}/N_b^{tot}=5$, 
$N_b^{tot}=100$ and $N_s^{tot}=0$ (corresponding to the isentropic expansion 
trajectory $S^{tot}/N_b^{tot}=5$ in Fig.~3(a)).
($B^{1/4}=180$~MeV, $m_s=150$~MeV)}
\label{fig4}
\end{figure}

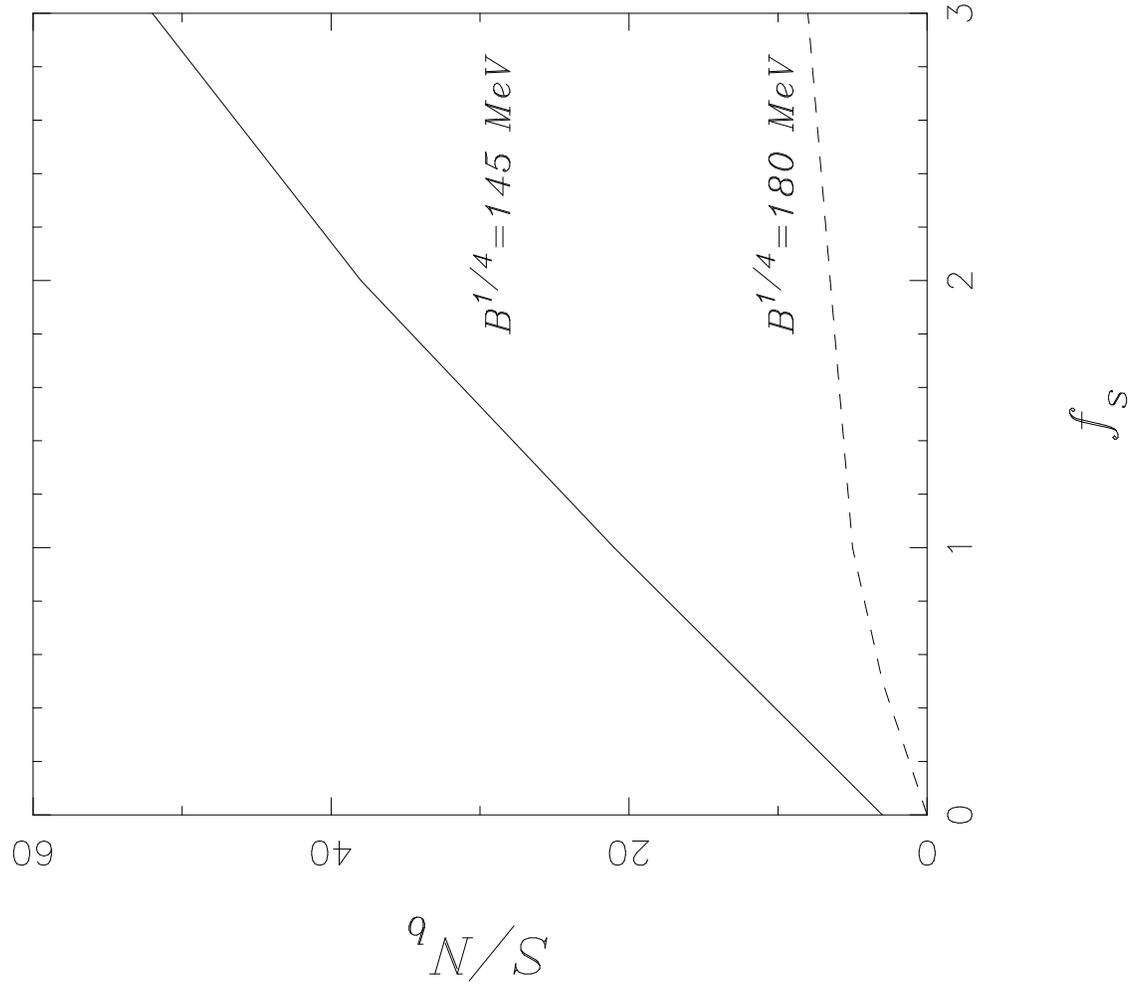
\begin{figure}
\caption{Maximum total entropy per baryon $S^{tot}/N_b^{tot}$ permitted by the 
formation of cold strangelets as a function of the total strangeness fraction 
$f_s^{tot}$ for a system with $N_b^{tot}=100$ and 
two values of bag constant $B$.} 
\label{fig5}
\end{figure}

\begin{references}

\bibitem{wit84}E. Witten, Phys. Rev. D {\bf 30}, 272 (1984).

\bibitem{liu84}H.-C. Liu and G. L. Shaw, Phys. Rev. D {\bf 30}, 1137 (1984).

\bibitem{gre87}C. Greiner, P. Koch and H. St\"{o}cker, Phys. Rev. Lett. 
{\bf 58}, 1825 (1987); C. Greiner, D. H. Rischke, H. St\"ocker and P. Koch,
Phys. Rev. D {\bf 38}, 2797 (1988); C. Greiner, P. Koch and H. St\"ocker,
Phys. Rev. D {\bf 44}, 3517 (1991).

\bibitem{bar90}See, e.g., J. Barrette {\it et al.}, Phys. Lett. 
B {\bf 252}, 550 (1990);
Phys. Rev. Lett. {\bf 70}, 1763 (1993); A. Aoki {\it et al.}, Phys. Rev. Lett.
{\bf 69}, 2345 (1992); K. Borer {\it et al.}, Phys. Rev. Lett. {\bf 72},
1415 (1994). For a recent review see B. S. Kumar, Nucl. Phys. {\bf A590},
29c (1995).

\bibitem{lee93}K. S. Lee and U. Heinz, Phys. Rev. D {\bf 47}, 2068 (1993).

\bibitem{barz90}H. W. Barz, B. L. Friman, J. Knoll and H. Schulz,
Phys. Lett. B {\bf 242}, 328 (1990).

\bibitem{mad93}J. Madsen, Phys. Rev. Lett. {\bf 70}, 391 (1993);
Phys. Rev. D {\bf 47}, 5156 (1993);
{\it ibid.} {\bf 50}, 3328 (1994).

\bibitem{he96}Y. B. He, C. S. Gao, X. Q. Li and W. Q. Chao, Phys. Rev. C 
{\bf 53}, 1903 (1996).

\bibitem{far84}E. Farhi and R.L. Jaffe, Phys. Rev. D {\bf 30}, 2379 (1984).

\bibitem{gil93}E. P. Gilson and R. L. Jaffe, Phys. Rev. Lett. {\bf 71},
332 (1993).

\bibitem{bal70}R. Balian and C. Bloch, Ann. Phys. {\bf 60}, 401 (1970);
T.H. Hansson and R.L. Jaffe, Ann. Phys. {\bf 151}, 204 (1983).

\bibitem{jen95}D. M. Jensen and J. Madsen, Proc. Strangeness and Quark 
Matter, Ed. G. Vassiliadis, A.D. Panagiotou, S. Kumar, and J. Madsen,
World Scientific (1995), p.220.

\bibitem{ber87}M. S. Berger and R. L. Jaffe, Phys. Rev. C {\bf 35}, 213 (1987);
{\bf 44}, 566 (E) (1991).

\bibitem{bal70a}R. Balian and C. Bloch, Ann. Phys. {\bf 64}, 271 (1970).

\bibitem{elz86}H.-T. Elze and W. Greiner, Phys. Lett. B {\bf 179}, 385 (1986).

\bibitem{hei86}U. Heinz, P. R. Subramanian, H. St\"ocker and W. Greiner,
J. Phys. G {\bf 12}, 1237 (1986).

\bibitem{hag80}R. Hagedorn and J. Rafelski, Phys. Lett. B {\bf 97}, 180 (1980).

\bibitem{surface}For calculations of surface tension see, {\it e.g.},
Z. Frei and A. Patk\'os, Phys. Lett. B {\bf 222}, 469 (1989); {\it ibid.} 
{\bf 247}, 381 (1990); 
B. Grossmann and M. L. Laursen, Nucl. Phys. {\bf B408}, 
637 (1993);
A. A. Coley and T. Trappenberg, Phys. Rev. D {\bf 50}, 4881 (1994); 
Y. Aoki and K. Kanaya, Phys. Rev. D {\bf 50}, 6921 (1994).

\bibitem{mac75}F. D. Mackie, Nucl. Phys. {\bf A245}, 61 (1975).

\bibitem{mar91}I. Mardor and B. Svetitsky, Phys. Rev. D {\bf 44}, 878 (1991).

\bibitem{sto80}H. St\"ocker, G. Graebner, J. A. Maruhn and W. Greiner,
Phys. Lett. B {\bf 95}, 192 (1980); H. St\"ocker and W. Greiner, Phys. Rep.
{\bf 137}, 277 (1986).

\bibitem{hei87}U. Heinz, K. S. Lee and M. J. Rhoades-Brown, Mod. Phys. Lett.
A {\bf 2}, 153 (1987).

\end{references}
\end{document}